# Comparision of Traditional and Fuzzy Failure Mode and Effects Analysis for Smart Grid Electrical Distribution Systems

Shravan Kumar Akula
School of Electrical Engineering and
Computer Sciences
University of North Dakota
Grand Forks, ND
shravankumar.akula@und.edu

Hossein Salehfar
School of Electrical Engineering and
Computer Sciences
University of North Dakota
Grand Forks, ND
h.salehfar@und.edu

Shayan Behzadirafi
Research and Development Department
New York Power Authority
New York, NY
shbehzadi@gmail.com

*Abstract*— Reliability Assessment is an indispensable technology for identifying, interpreting, and lessening the potential failures in safety-critical systems like smart grids. Failure modes and effects analysis (FMEA) is one of the well-documented techniques for risk analysis to study the impact of failure modes on safety critical systems like smart grid. In traditional FMEA failure modes are prioritized based on a numeric assessment known as risk priority number. Risk priority number (RPN) is defined as the product of three risk factors namely severity (S), occurrence (O), and detection (D). These risk factors are generally attained by extensive team efforts and judgments which can lead to errors and inconsistencies. To address the shortcomings of the traditional FMEA, a fuzzy-based FMEA approach is proposed to generate reliable risk priority rankings. In this study, traditional and fuzzy-based FMEA risk priority rankings for smart grid electrical distribution systems are compared and recommendations are made based on the analysis. Results prove the efficiency of the proposed fuzzy-FMEA method.

## I. INTRODUCTION

It is undeniable that renewable energy sources are the future of the modern power grid. Large-scale deployments of renewable energy sources will have impacts on the grid planning and operation due to the concerns over availability and reliability [1]. Increased installations of renewables increase system complexity, if not planned properly can lead to a decline in grids performance and in the worst-case blackouts.

To quantify the level of uncertainty in smart grid, reliability and risk assessment is the ideal choice. These assessments systematically identify and calculate the probability of failure sequences that will lead to the grid failures, and collect and document the basic events.

Usage of reliability analysis tools in smart grids helps operators to develop prevention and maintenance strategies that will create a reliable grid with minimal installation and maintenance costs [2]. Reliability Centered Maintenance (RCM) which is defined as "a process used to determine the maintenance requirements of any physical asset in operating context" [3] is the perfect choice for smart grids. When applied correctly, RCM will quickly show astonishing improvements in maintenance efforts. For example, an RCM based program for Turkish National Power Systems is presented in [4] where the system is decomposed into individual blocks and the failure modes for each block are individually studied to attain an optimized maintenance schedule for the transmission system. FMEA complements RCM perfectly. With a thorough FMEA analysis, smart grid owners can understand their assets better and implement strategies that address failure modes.

The rest of the paper is organized as follows. The literature review is given in section II, FMEA and fuzzy FMEA are discussed in section III, results for the fuzzy FMEA are documented in section IV and the paper ends with a conclusion and future works in section V.

## II. LITERATURE REVIEW

There are two categories of reliability and risk evaluation techniques: analytical (e.g. fault tree analysis) and simulation methods (Monte Carlo simulation) [5]. Based on the objective, a proper evaluation method must be selected. The necessity for the reliability assessment in the field of power systems was identified decades ago. Reliability, availability, and maintainability analysis for a photovoltaic system was proposed in [6] but the lack of field data leaves gaps in research. Using the Monte Carlo simulation method, the reliability of the time-varying loads is performed in [7] using a load duration curve of the system. But simulation methods are computationally expensive as a large number of variables are bounded to different constraints.

FMEA is a comprehensive reliability analysis tool that requires minimal computational power [8]. Based on the photovoltaic system located at Northeast Solar Energy Research Center and Brookhaven National laboratory a FMEA analysis for the systems components and sub-components was proposed in [9].

To obtain an optimized maintenance schedule, develop a comprehensive database of the maintenance requirements, and to identify and prolong the life of critical components of microgrids, authors of this work have chosen the RCM approach. Fault tree analysis was used for the reliability assessment of the microgrid with photovoltaics, wind turbines, and battery energy storage systems as distributed generators [8]. Several importance measures were proposed to help the grid operators to improve the operating performance, and to enhance reliability. Modern microgrids consist of complex communication infrastructure with devices such as controllers, actuators, and sensors that are prone to cyber-attacks. To address this, risk analysis of cyber components performed in [10] using the traditional FMEA, and risk matrices were proposed to identify the critical cyber components in the smart grid.

Traditional FMEA has several drawbacks which are discussed in section III. To improve the efficiency of the traditional FMEA based risk analysis and to provide a mathematical basis for the risk assessment, authors are proposing a fuzzy-based FMEA in this work. To test the efficacy of the proposed Fuzzy-FMEA, authors have tested the proposed method with the smart electrical distribution system proposed in [11], and have compared the traditional- and fuzzy-based risk priority numbers (RPN) of that system.

There are numerous reliability studies that focus on different methods with a wide range of indices. A single method cannot serve all the aspects of reliability and risk assessment. Authors have conducted an extensive literature review and found out that there is little to no work in the area of FMEA for smart grids. To fill the gaps in the research authors are proposing a fuzzy based FMEA (Fuzzy-FEMA) in this paper, the advantages of which are explained in the next section.

### III. FAILURE MODE AND EFFECTS ANALYSIS

#### A. Traditional FMEA

Introduced in the 1940s by the US military, FMEA is a systematic methodology that identifies the potential failure modes, their causes, and consequences on the system performance. These methods identify the critical areas for process improvements by prioritizing the highest-priority failure modes. The procedure for traditional FMEA is shown in Fig. 1.

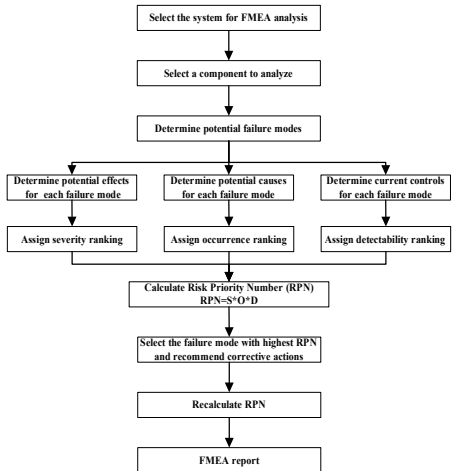

Fig. 1. Traditional FMEA procedure

For evaluating risk and to prioritize the failure modes, risk priority numbers (RPN) are used. RPN is defined as the product of three risk factors, namely severity (S, severity of the failure mode), occurrence (O, the likelihood of failure), and detection (D, likelihood that failure will be detected with present controls).

$$RPN = S*O*D \quad (1)$$

Risk factors S, O, and D are rated on a scale ranging from 1 to 10. The ranking system typically used in traditional FMEA is provided in [12]. A high RPN implies a greater risk on systems operation. Although traditional FMEA is a proactive risk assessment tool it has several shortcomings [13] some of which are as follows:

- Relation between different components is not considered
- Risk factors are treated with equal weights, i.e., their relative importance is ignored
- Calculation of RPN lacks scientific basis
- Different combinations of the risk factors S, O, and D can lead to the same RPN which in some cases can lead to ignoring high-risk failure modes

To address the shortcoming in the traditional FMEA authors in this work are proposing a fuzzy logic based FMEA (Fuzzy-FEMA) method and show it is a more effective tool for characterizing the system reliability, compared to the traditional FEMA based methods.

#### B. Fuzzy-FMEA

Fuzzy logic considers inaccuracy, subjectivity, uncertainty, and imprecision in decision making, so it is a perfect choice for risk assessment. As the failure events cannot be described only in numbers, precise determination of risk factors S, O, and D with linguistic variables makes fuzzy logic risk assessment qualitative and operational. A fuzzy inference system (FIS) is a popular computational method for formulating the relation among input and output data based on the concepts of fuzzy set Failure Mode and Effects Analysis (FMEA) is a tool used to identify potential failure modes in a system and to assess the associated risks. Reliability-Centred Maintenance (RCM) is a tool used to identify the maintenance requirements of a system in order to ensure its reliability.theory. The architecture for the FIS is shown in Fig. 2.

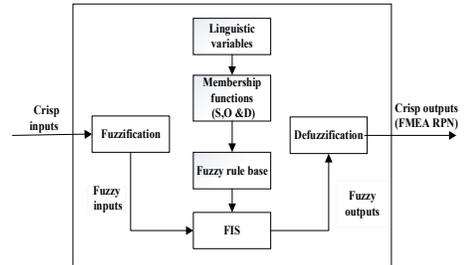

Fig. 2. FIS system

The process of converting the crisp inputs to a fuzzy set is referred to as fuzzification. FIS risk assessment model is implemented using the MATLAB fuzzy toolbox. In this work, Mamdani FIS was used as it is intuitive, has a more interpretable rule base, and widespread acceptance. For the S,

O, and D input fuzzy logic membership functions, the triangular membership function is used as the straight line membership functions have the advantage of simplicity, speed, and computational efficiency [14] as shown below.

$$triangle(x; a, b, c) = \begin{cases} 0, & x \leq a \\ \frac{x-a}{b-a}, & a \leq x \geq b \\ \frac{c-x}{c-b}, & b \leq x \leq c \\ 0, & c \leq x \end{cases} \quad (2)$$

where a, b, and c (a<b<c) are the coordinates of the three corners for the membership function.

The Gaussian membership function is used for the FIS FEMA-RPN output for its smoothness and concise notation [14].

$$gaussian(x, c, \sigma) = e^{-\frac{1}{2}\left(\frac{x-c}{\sigma}\right)^2} \quad (3)$$

where 'c' represents the center and '$\sigma$' is the width of the membership function.

Based on the defined input and output parameters, a relation between each parameter was identified in the form of 'if-then' rules from type 'and'. A total of 125 rules were recorded.

For the defuzzification process, centroid defuzzification was used. This method is the most commonly used defuzzification technique as it is known for its accuracy but has more computation time.

$$defuzzification = \frac{\int_a^b x * \mu_A(X) dx}{\int_a^b \mu_A(X) dx} \quad (4)$$

Where 'x' represents the sample element and '$\mu_A(X)$' is the membership function on the interval [a b].

## IV. METHODOLOGY

### A. Fuzzy Inference System (FIS) Model

A fuzzy inference system (FIS) is a type of artificial intelligence system that uses fuzzy logic to make decisions. Fuzzy logic is a way of representing knowledge that is not precise. In a fuzzy inference system, each input is given a fuzzy value, and the system calculates a fuzzy output. Mamdani [15] and Sugeno [16] are the two popular fuzzy reasoning systems. Mamdani inference systems are more versatile and accurate than Sugeno inference systems. Mamdani systems can handle more complex input data, has widespread acceptance, and produce more accurate results. Sugeno systems are better suited for simpler input data, and they are less likely to produce accurate results for complex systems. In this work, using the Mamdani method and the 'Fuzzy Logic designer' app in MATLAB, the proposed FIS was developed as shown in Fig.3.

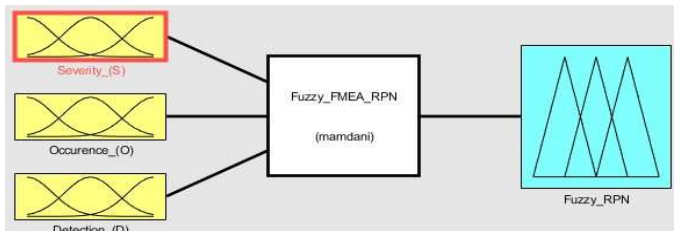

Fig. 4. Fuzzy Inference System structure

Inputs *S* and *O* have five fuzzy sets (Very Low, Low, Moderate, High, and Very High) represented by the Triangular membership functions. Input *D* has five fuzzy sets (Very High, High, Moderate, Low, and Very Low) represented by the triangular membership function. Output function *Fuzzy_RPN* has five fuzzy sets (Very Low, Low, Medium, High, and Very High) represented by the Gaussian membership function. Figs. 4 and 5 illustrate the input and output membership functions.

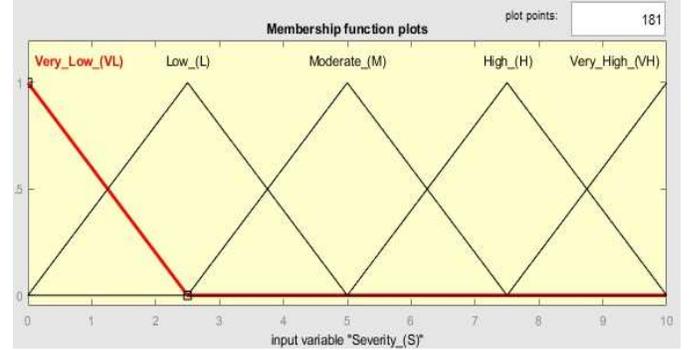

Fig, 4. Input membership functions

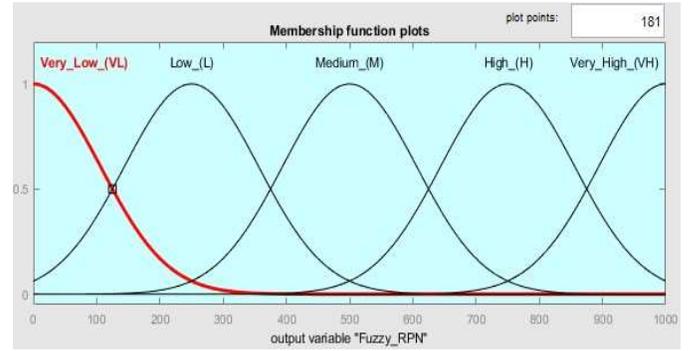

Fig. 5. Fuzzy FMEA risk membership functions

Using feedback from experts and the MATLAB rule editor 125 "If then" rules of fuzzy conditionals statements were defined. These rules compare the input variables with the membership functions to obtain the values of linguistic inputs, combine the membership values depending on the weightage of each rule, and generate and aggregate the qualified conditional statements to produce a crisp output. For the defuzzification of fuzzy outputs, authors chose the *Centroid* method as it produces consistent results while taking the height and width of the fuzzy output into account [17]. Using the centroid method (eq. 4.) for defuzzification, numerical value linguistic outputs are calculated as shown in Fig. 6.

Relation between the input membership functions namely severity (S), occurrence (O), detection (D), and output membership function fuzzy RPN can be represented by a 3D responding function (Fig. 7.). When comparing the fuzzy-FMEA approach to the classical FMEA method, it is clear that the fuzzy logic method offers a larger range of risk assessment and narrower gaps between different levels of risk, implying more accuracy.

The presented model differs from earlier techniques in that it performs a pre-assessment of the input variables by establishing evaluation criteria and generating fuzzy sets for them.

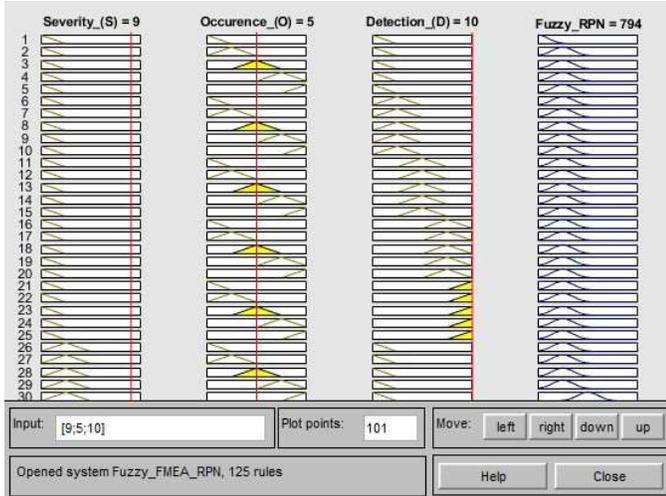

Fig.6. Sample rule viewer

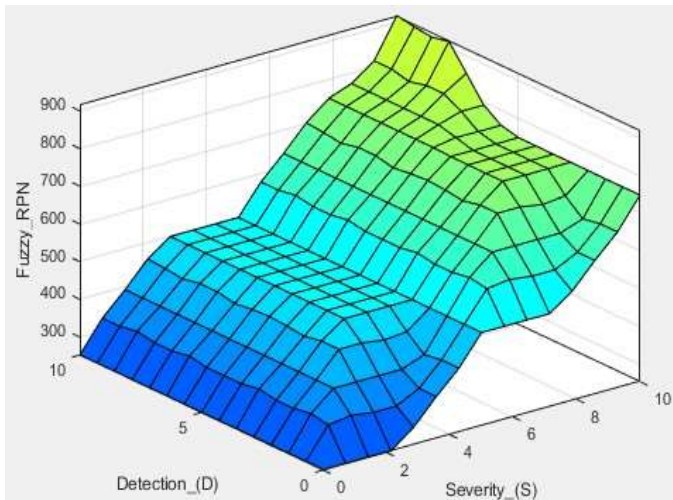

Fig. 7. The surface viewer of the fuzzy logic system

## B. Results and Discussion

After building the Fuzzy-FMEA model, authors tested and analyzed the FMEA for smart grid distribution systems in [2] to evaluate the risk analysis for power and cyber network main components. In the traditional FMEA, Servers (S=8, O=6, D=10, and RPN=480) and transformers (S=9, O=5, D=10, and RPN=450) are determined to be the most critical components whose failures can compromise the operation of the smart grid. Human Machine Interfaces, Switches, and Intelligent electronic devices are the critical components in cyber equipment. Ethernet and optical fiber net links, energy boxes (Energy management systems) have the lowest RPN of all components due to their low failure and repair rates.

Despite the enormous consequences cyberattacks may have on a system, failure modes related to cyberattacks are not considered as high-risk failure modes in [2]. This explains the low occurrence ratings despite the predicted rise in the number of cyberattacks.

Events, intelligent electronic devices with communication failure mode (S=6, O=5, D=8, and RPN-240), transformer with winding breakdown mode (S=6, O=4, D=10, and RPN=240), server with power outage breakdown mode (S=7, O=3, D=10, and RPN=210), and circuit breaker with insulation breakdown mode (S=6, O=5, D=7, and RPN=210), among others, have the same RPN value despite different combinations of S, O and D values. The degree of risk for the individual components and failure modes is not taken into account. Another issue with the traditional RPN ranking is that it disregards the qualified relevance of S, O, and D. Consequently, these three parameters are considered to have similar outcomes. However, in actual reality qualified importance measures of system components exist that should be recognized and accounted for. Fuzzy-FMEA addresses the issues mentioned above by prioritizing risks better in the case of the same RPN as it takes the influence of individual variables (S, O, and D) into account.

To address the shortcomings of the traditional FMEA authors have used the developed Fuzzy-FMEA model and the results are tabulated in table-I. A total of 55 equipment failure modes are identified and Fuzzy-RPN is calculated.

TABLE I. TRADITIONAL AND FUZZY FMEA RPN RESULTS (T= TRADITIONAL , F=FUZZY)

| Component | Failure mode | T-RPN | F-RPN | T-Rank | Fuzzy RPN Rank |
|---|---|---|---|---|---|
| Server | Hardware crash | 480 | 753 | 1 | 2 |
| Transformer | Explosion | 450 | 794 | 2 | 1 |
| Human Machine Interface | Operational failure | 400 | 747 | 3 | 9 |
| Intelligent electronic device | Control failure | 392 | 752 | 4 | 7 |
| Bus bar | Structural integrity loss | 378 | 677 | 5 | 15 |
| Cable | Operational failure | 360 | 602 | 6 | 22 |
| Switch | Operational failure | 360 | 602 | 7 | 23 |
| Bus bar | electrical loss continuity | 320 | 753 | 8 | 3 |
| Bus bar | electrical disturbances | 320 | 753 | 9 | 4 |
| Transformer | Winding displacement | 315 | 677 | 10 | 16 |
| Circuit breaker | Busing breakdown | 300 | 602 | 11 | 24 |
| Server | Data errors | 300 | 602 | 12 | 25 |
| Transformer | Winding overheating | 294 | 677 | 13 | 17 |
| Cable | Manufacturing defect | 280 | 747 | 14 | 10 |
| Circuit breaker | Degradation of contacts | 270 | 602 | 15 | 26 |
| Switch | Performance degradation | 252 | 602 | 16 | 27 |
| Intelligent electronic device | Communication failure | 240 | 602 | 17 | 28 |
| Transformer | Winding degradation | 240 | 602 | 18 | 29 |
| Transformer | Breakdown of bushings | 240 | 753 | 19 | 5 |

| Transformer | Rupture in oil tank | 216 | 686 | 20 | 12 |
| Intelligent electronic device | Power outage | 210 | 686 | 21 | 13 |
| Server | Power outage | 210 | 602 | 22 | 30 |
| Circuit breaker | Failure in insulation | 210 | 747 | 23 | 11 |
| Server | Physical security failure | 200 | 602 | 24 | 31 |
| Circuit breaker | Hot spot in bushing terminals | 192 | 753 | 25 | 6 |

The final Fuzzy-FMEA analysis results for different component failure modes are presented in Table I in which twenty-five failure modes with the highest traditional-RPN and their corresponding Fuzzy-FMEA PRN are documented. Failure modes are ranked according to their RPN.

From the results transformers failure modes with rankings 1, 5, 9, 12, 16, and 29, servers, and human-machine interfaces with their respective failure modes are the most vital components in the system which when failed can compromise the operation of the smart grid. Bus bars with different failure modes ranked at 3, 4, 15, and 21 are critical components and when failed they can cause considerable loss of service and severe system disruptions. Similar to the traditional FMEA ethernet and optical fiber links are found to be the most reliable components in the system.

As can be seen, the proposed Fuzzy-FMEA model outcomes are far more realistic than those from the traditional FMEA. The most notable difference is that fuzzy FMEA uses fuzzy logic to account for the uncertainty in predicting failure modes where there is insufficient data or knowledge to make precise decisions. This is done by assigning weights to each input which allows to prioritize mitigation efforts. Finally, fuzzy-FMEA also provides a means for dealing with multiple experts with conflicting opinions. This is done by aggregating the opinions of the experts using fuzzy set theory. This results in a more robust and accurate analysis than would be possible with traditional FMEA. Also, the Traditional FMEA model takes human judgment into account when making decisions. Whereas in Fuzzy-FMEA human partisanship is eliminated by combining numerical and linguistic variables to evaluate the subjectiveness. For example:

- The value of T-RPN for component database server with a power outage failure mode is 210 (S=7, O=3, and D=10) and the value of F-RPN 602.
- The value of T-RPN for a circuit breaker with insulation failure is 210 (S=6, O=5, and D=7) is 210 and the value of F-RPN is 747.

Based on the traditional FMEA model the RPN values are the same. The traditional FMEA model has missed the importance of the value of parameter *Detection*. Traditional FMEA does not consider the weight of the individual input parameters, but the developed fuzzy model provided the correct weight for the parameter and ranked the failure modes correctly.

And,

- Transformer with cooling system failure mode has a T-RPN of 168 (S=8, O=3, and D=8) and F-RPN of 752.
- T-RPN value for switch with physical security failure mode is 80 from the multiplication of three parameters (S=8, O=8, and D=5) but has a value of F-RPN=686.

In the traditional FMEA based maintenance plan, neither of the failure modes analyzed above would be considered because their RPNs are below 200. But by applying the fuzzy logic it became clear that the influence of input risk factor variables had a considerable impact on risk prioritizing.

Results prove that the proposed Fuzzy-FMEA model is resilient, sensitive to tiny changes in input parameters, and has a major feature of a reaction to quantifiable risk, in contrast to standard RPN, which uses a scale of 1 to 1000 and relies only on RPN with the possibility of high-risk failure modes excluded.

The RPN plots of traditional and Fuzzy FMEA are shown in Fig. 8. The failure modes with the similar F-RPN score in the plot are mainly determined by the high *Severity* and *Detection* ratings) and the fuzzy rule-base.

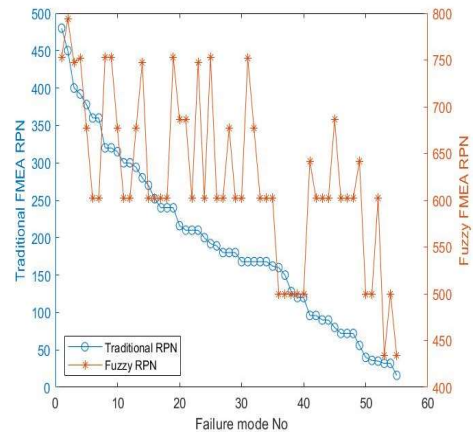

Fig. 8. Traditional and Fuzzy FMEA RPN vs Failure mode No

The plots of traditional and Fuzzy-FMEA rankings are shown in Fig. 9. For most of the failure modes, F-RPN rankings followed a trend similar to traditional FMEA. The model identified vulnerabilities in the smart-grid and ranked them in the correct order.

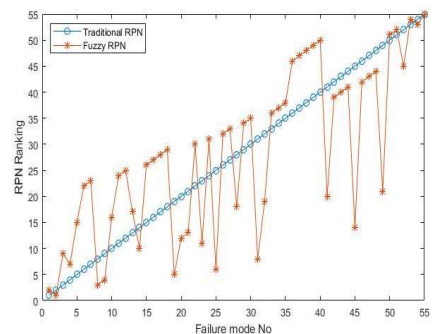

Fig. 9. Traditional and Fuzzy-FMEA failure mode rankings

Human interference in smart grids must be considered as a part of reliability and risk analysis as they can play a significant role in future smart grids. The report published on blackouts in the United States and Canada shows the lack of monitoring can lead to cascading failures in the system [18]. An unmonitored transformer fire and explosion resulted in multiple line tripping's and a shutdown of the power plants [19]. These accidents are attributed to human errors which proves the necessity for the inclusion of human errors in risk analysis. Although some human errors are included as a part of the present risk analysis, the impact of human errors on emergency dispatch, maintenance, and operation scenarios is not included. These issues will be addressed in future work of the authors.

Finally, in this work economic constraints resulting from failure modes are not considered. In the future authors intend to develop an economic impact based FMEA that prioritizes risks based on economic impact value.

## V. Conclusion and Future work

This work proposes a novel approach to determine distribution system component failure modes by evaluating their Fuzzy-FEMA based RPN values and to identify the root causes of their failure. Traditional FMEA RPN values were compared with the Fuzzy FMMA based RPN values to address the drawbacks of the former method. Accurate and consistent results prove that the three input parameters used to create linguistic membership functions are an effective way for risk analysis using FMEA. A fuzzy conclusion prevents the appearance of identical RPN values for distinct risk factor sets, and it is possible to define priorities in rating different defects and their impact on specific corrective steps taken based on the results.

In terms of future research, finding a method based on fuzzy logic that allows for the testing of parameters incorporated in the fuzzy decision-making mechanism to assess and rank risks associated with the type of defect that could occur in the system's operation is critical. When certain FMEA corrective measures are used, fuzzy logic is an effective methodology in the analysis of priority rankings, having an impact on the reliability system as well. In addition to Fuzzy-FMEA authors plan to include an artificial neural network model for FMEA evaluation to conduct an effective FMEA process to help FEMA working teams minimize their workload, save design time, and assure system success.